\documentclass[aps,preprint]{revtex4}%
\usepackage{amsfonts}
\usepackage{amsmath}
\usepackage{amssymb}
\usepackage{graphicx}%
\providecommand{\U}[1]{\protect\rule{.1in}{.1in}}

\begin{document}
\title{The non-Hermitian Swanson model with a time-dependent metric}
\author{Andreas Fring$^{\bullet}$ and Miled H. Y. Moussa$^{\bullet,\circ}$}
\affiliation{$^{\bullet}$Department of Mathematics, City University London, Northampton Square,
London EC1V 0HB, UK}
\affiliation{$^{\circ}$Instituto de F\'{\i}sica de S\~{a}o Carlos, Universidade de S\~{a}o
Paulo, P.O. Box 369, S\~{a}o Carlos, 13560-970, SP, Brazil}

\begin{abstract}
We provide further non-trivial solutions to the recently proposed
time-dependent Dyson and quasi-Hermiticity relation. Here we solve them for
the generalized version of the non-Hermitian Swanson Hamiltonian with
time-dependent coefficients. We construct time-dependent solutions by
employing the Lewis-Riesenfeld method of invariants and discuss concrete
physical applications of our results.

\end{abstract}

\maketitle

\section{Introduction}

PT symmetric (PTS) quantum mechanics has attracted increasing attention since
is was demonstrated that PTS Hamiltonians possess real spectra \cite{BB1998}
and allow for a unitary evolution with a redefined inner product
\cite{Geyer,Mostafa}. Phase transitions between the regimes of unbroken and
broken PT symmetry, which are a key feature in the energy spectrum are well
understood to occur when two real eigenvalues coalesce to form complex
conjugate pair \cite{BB1998}. Many interesting new results have recently
emerged from the application of PTS concepts to different areas of physics, in
the classical and the quantum domain, on both fronts, theoretical as well as
experimental. We mention here a few, such as the design of an
ultralow-threshold phonon laser \cite{PL}, the demonstration of defect states
\cite{OL} and beam dynamics \cite{BD} in PTS optical lattices, and the fact
that the Jarzynski equality generalizes to PTS domain \cite{J}. Reinforcing
the practical features, there are optical structures described by PTS concepts
that enable unprecedented control of light \cite{CL}. At a classical level,
PTS properties have also been observed in a variety of experimental set-ups,
ranging from quantum optics \cite{QO} to NMR \cite{NMR} and superconductivity
\cite{S}.

Although the grounds for treating non-Hermitian Hamiltonians using
time-independent metric operators have been extensively studied and well
established \cite{CB,CA}, the generalization to time-dependent (TD) metric
operators has raised controversy \cite{M,Z,Wang,MM}. In Ref. \cite{M},
Mostafazadeh has demonstrated that using a TD metric operator one can not
ensure the unitarity of the time-evolution simultaneously with the
observability of the Hamiltonian. From this perspective, with which we agree,
the authors of Refs. \cite{Z,Wang,MM} fail to ensure a unitary time-evolution
by insisting on the observability of the Hamiltonian. However, we have
recently suggested \cite{Andreas} that this is not an obstacle and certainly
not a no-go theorem. It is very common in the context of PTS quantum mechanics
that certain operators, such as position or momentum, may become
non-observable auxiliary variables and only their quasi-Hermitian counterparts
can be measured. In \cite{Andreas} we take the view that the Hamiltonian,
meaning the operator that satisfies the TD Schr\"{o}dinger equation (SE),
joins this set of observables in the scenario where a TD metric operator is
considered. For this proposal to be meaningful the TD quasi-Hermiticity
relation and TD dyson relation need to possess non-trivial solutions. When
this is the case, we have unitary time evolution and well defined observables.

Here we provide new non-trivial solution to this set of equations for
generalized time-dependent version of Swanson Hamiltonian \cite{Swanson} by
solving its TDSE and by computing some observables. In order to solve the SE,
we shall adapt a method presented in Ref. \cite{MBS,SanjibA} for treating TD
Hermitian Hamiltonians. This method takes advantage of a unitary TD
transformation on the SE, here replaced by a non-unitary transformation to
conform with non-Hermitian Hamiltonians, and the diagonalization of a TD
Invariant on the Lewis and Riesenfeld framework \cite{LR}.

The authors in Ref. \cite{MBS} pursued the solution of the SE governed by a
general TD quadratic Hamiltonian in order to investigate the mechanism of
squeezed states following from the nonlinear amplification terms of the
Hamiltonian \cite{Walls,Scully}. Here, we shall focus on the technique to
treat a TD non-Hermitian Hamiltonian, leaving open the possibility of further
analysis of the squeezing mechanism coming from the nonlinear terms of a TD
non-Hermitian Hamiltonian.

\section{Non-Hermitian Hamiltonian systems with TD metric}

Let us briefly review the scheme proposed in \cite{Andreas}: We consider a
non-Hermitian TD Hamiltonian $H(t)$ whose associated SE, $i\partial
_{t}\left\vert \psi(t)\right\rangle =H(t)\left\vert \psi(t)\right\rangle $, is
mapped, by means of the Hermitian TD operator $\eta(t)$, into the SE
$i\partial_{t}\left\vert \phi(t)\right\rangle =h(t)\left\vert \phi
(t)\right\rangle $, where the corresponding wave functions are transformed as
$\left\vert \phi(t)\right\rangle =\eta(t)\left\vert \psi(t)\right\rangle $ and
the Hamiltonians are related by means of the \textit{TD Dyson relation}
\begin{equation}
h(t)=\eta(t)H(t)\eta^{-1}(t)+i\left[  \partial_{t}\eta(t)\right]  \eta
^{-1}(t)\text{.} \label{1}%
\end{equation}
We set here $\hbar=1$. The key feature in this equation is the fact that
$H(t)$ is no longer quasi-Hermitian, i.e. related to $h(t)$ by means of a
similarity transformation, due to the presence of the last term. Thus $H(t)$
is not a self-adjoined operator and therefore not observable. Using the
Hermiticity of $h(t)$ we then derived the \textit{TD quasi-Hermiticity
relation}
\begin{equation}
H^{\dag}(t)\rho(t)-\rho(t)H(t)=i\partial_{t}\rho(t), \qquad\rho(t)=\eta
^{\dagger}\eta, \label{2}%
\end{equation}
replacing the standard quasi-Hermiticity relation for a time-independent
$\rho$, given by $H^{\dag}\rho=\rho H$. In fact, the TD quasi-Hermiticity
relation ensures the TD probability densities in the Hermitian and
non-Hermitian systems to be related in the standard form
\begin{equation}
\left\langle \psi(t)\left\vert \tilde{\psi}(t)\right.  \right\rangle _{\rho
}=\left\langle \psi(t)\left\vert \rho(t)\right\vert \tilde{\psi}%
(t)\right\rangle =\left\langle \phi(t)\left\vert \tilde{\phi}(t)\right.
\right\rangle \text{.} \label{3}%
\end{equation}

With the assumption that $\rho(t)$ is a positive-definite operator, it plays
the role of the TD metric and we conclude that any self-adjoined operator
$o(t)$, i.e. observable, in the Hermitian system possesses a counterpart
$O(t)$ in the non-Hermitian system given by
\begin{equation}
O(t)=\eta^{-1}(t)o(t)\eta(t)\text{,} \label{4}%
\end{equation}
in complete analogy to the time-independent scenario. Thus as long as the
generalized equations (\ref{1}) and (\ref{2}) posses non-trivial solutions for
$\eta(t)$ and $\rho(t)$, respectively, we have a well defined physical system
with TB observables and unitary time-evolution governed by a TD non-Hermitian
Hamiltonian. Albeit we have the slightly unusual feature that the TD
Hamiltonian $H(t)$ does not belong to the set of observables.

\section{The generalized time-dependent Swanson Hamiltonian}

The system we wish to investigate here is related to the non-Hermitian TD
Swanson Hamiltonian%
\begin{equation}
H(t)=\omega(t)\left(  a^{\dagger}a+1/2\right)  +\alpha(t)a^{2}+\beta
(t)a^{\dagger2}\text{,} \label{5}%
\end{equation}
where $a$ and $a^{\dagger}$ are bosonic annihilation and creation operators,
for instance of a light field mode. In comparison with time-independent case
all parameters have acquired an explicit time-dependence $\omega
(t),\alpha(t),\beta(t) \in\mathbb{C}$. Clearly when $\omega(t) \notin
\mathbb{R}$ or $\alpha(t)\neq\beta^{\ast}(t)$ the Hamiltonian (\ref{5}) is not
Hermitian. It becomes PT-symmetric when demanding $\omega(t),\alpha
(t),\beta(t)$ to be even functions in $t$ or generic functions of $it$.

Let us now solve the TD Dyson equation by making the following general and for
simplicity Hermitian Ansatz for the Dyson map
\begin{align}
\eta(a,a^{\dagger},t)  &  =\exp\left[  \epsilon(t)\left(  a^{\dagger
}a+1/2\right)  +\mu(t)a^{2}+\mu^{\ast}(t)a^{\dagger2}\right] \label{6a}\\
&  =\exp\left[  \lambda_{+}(t)K_{+}\right]  \exp\left[  \ln\lambda_{0}(t)
K_{0}\right]  \exp\left[  \lambda_{-}(t)K_{-}\right]  \text{.} \label{6}%
\end{align}
We require here the variant (\ref{6}) of our Ansatz to be able to compute the
time-derivatives of $\eta$. The equality follows by recalling that
$K_{+}=a^{\dagger2}/2$, $K_{-}=a^{2}/2$, $K_{0}=(a^{\dagger}a/2+1/4)$ form an
$SU(1,1)$-algebra, such that the group element in (\ref{6a}) can be Iwasawa
decomposed according to \cite{Klimov}. The TD coefficients read
\begin{subequations}
\label{7}%
\begin{align}
\lambda_{+}  &  =\frac{2\mu^{\ast}\sinh\Xi}{\Xi\cosh\Xi-\epsilon\sinh\Xi
}\text{,}\label{7a}\\
\lambda_{-}  &  =\lambda_{+}^{\ast}\text{,}\label{7b}\\
\lambda_{0}  &  =\left(  \cosh\Xi-\frac{\epsilon}{\Xi}\sinh\Xi\right)
^{-2}\text{,} \label{7c}%
\end{align}
where we abbreviated the argument of the hyperbolic functions to $\Xi
=\sqrt{\epsilon^{2}-4\left\vert \mu\right\vert ^{2}}$, demanding $\epsilon$ to
be real and $\epsilon^{2}-4\left\vert \mu\right\vert ^{2}\geq0$.

The notation may be simplified even further when introducing some new
quantities. Similarly as in \cite{Musumbu} we define $z=2\mu/\epsilon
=\left\vert z\right\vert e^{i\varphi}$ within the unit circle, such that we
obtain $\Xi=\epsilon\sqrt{1-\left\vert z\right\vert ^{2}}$. Furthermore, we
define $\Phi=\left\vert z\right\vert /\Gamma_{-}$ with $\Gamma_{\pm}%
=1\pm\tilde{\Xi}\coth\Xi$, $\tilde{\Xi}=\Xi/\epsilon$, $\tilde{\Gamma}_{\pm
}=\Gamma_{\pm}/\tilde{\Xi}$, and finally $\chi=\tilde{\Gamma}_{+}%
/\tilde{\Gamma}_{-}=2/\Gamma_{-}-1=2\Phi/\left\vert z\right\vert -1$. The
notation settled, the coefficients in (\ref{7a})-(\ref{7c}) simplify to
\end{subequations}
\begin{subequations}
\label{8}%
\begin{align}
\lambda_{+}  &  =-\Phi e^{-i\varphi}\text{,}\label{8a}\\
\lambda_{-}  &  =-\Phi e^{i\varphi}\text{,}\label{8b}\\
\lambda_{0}  &  =\frac{1}{\tilde{\Gamma}_{-}^{2}\sinh^{2}\Xi}=\Phi^{2}%
-\chi\text{.}\label{8c}%
\end{align}
where $\sinh^{2}\Xi=\tilde{\Xi}^{2}\Phi^{2}/\left[  \left\vert z\right\vert
^{2}\left(  \Phi^{2}-\chi\right)  \right] $ $=\tilde{\Xi}^{2}\lambda
_{+}\lambda_{-}/\left\vert z\right\vert ^{2}\lambda_{0}$.

Using the relations
\end{subequations}
\begin{equation}
\eta(t)%
\begin{pmatrix}
a\\
a^{\dagger}%
\end{pmatrix}
\eta^{-1}(t)=\pm\frac{1}{\sqrt{\lambda_{0}}}%
\begin{pmatrix}
-1 & \lambda_{+}\\
-\lambda_{-} & \chi
\end{pmatrix}%
\begin{pmatrix}
a\\
a^{\dagger}%
\end{pmatrix}
\text{,} \label{9}%
\end{equation}
we obtain, after some algebra, the transformed Hamiltonian
\begin{align}
h(z,\epsilon,t)  &  =\eta(t)H(t)\eta^{-1}(t)+i\dot{\eta}(t)\eta^{-1}%
(t)\nonumber\\
&  =W(z,\epsilon,t)(a^{\dagger}a+1/2)+V(z,\epsilon,t)a^{2}+T(z,\epsilon
,t)a^{\dagger2}\text{,} \label{10}%
\end{align}
where the coefficient functions are
\begin{subequations}
\label{11}%
\begin{align}
W(z,\epsilon,t)  &  =-\frac{1}{\lambda_{0}}\left[  \omega\left(  \chi
+\lambda_{+}\lambda_{-}\right)  \right.  \left.  +2\left(  \alpha\lambda
_{+}+\beta\chi\lambda_{-}\right)  -\frac{i}{2}\left(  \dot{\lambda}%
_{0}-2\lambda_{+}\dot{\lambda}_{-}\right)  \right]  \text{,}\label{11a}\\
V(z,\epsilon,t)  &  =\frac{1}{\lambda_{0}}\left(  \alpha+\omega\lambda
_{-}+\beta\lambda_{-}^{2}+\frac{i}{2}\dot{\lambda}_{-}\right)  \text{,}%
\label{11b}\\
T(z,\epsilon,t)  &  =\frac{1}{\lambda_{0}}\left[  \omega\chi\lambda_{+}%
+\alpha\lambda_{+}^{2}+\beta\chi^{2}+\frac{i}{2}\left(  \lambda_{0}%
\dot{\lambda}_{+}+\lambda_{+}^{2}\dot{\lambda}_{-}-\lambda_{+}\dot{\lambda
}_{0}\right)  \right]  \text{.} \label{11c}%
\end{align}
As common the overhead dot denotes derivatives with respect to time.

For the Hamiltonian $h(t)$ to be Hermitian we need to impose $W$ to be real
and in addition $T=V^{\ast}$. From the first constraint we derive the equality%
\end{subequations}
\begin{equation}
\dot{\lambda}_{0}=2\left\vert \omega\right\vert \left(  \chi+\Phi^{2}\right)
\sin\varphi_{\omega}+2\Phi\left[  \dot{\Phi}+2\left\vert \alpha\right\vert
\sin\left(  \varphi-\varphi_{\alpha}\right)  -2\left\vert \beta\right\vert
\chi\sin\left(  \varphi+\varphi_{\beta}\right)  \right]  \text{,} \label{12}%
\end{equation}
while the second one leads to the coupled nonlinear differential equations
\begin{align}
\dot{\Phi}  &  =\frac{2}{\chi-1}\left\{  \left[  \left\vert \omega\right\vert
\Phi\sin\varphi_{\omega}+\left\vert \alpha\right\vert \sin\left(
\varphi-\varphi_{\alpha}\right)  \right]  \left(  1-\Phi^{2}\right)  \right.
\left.  +\left\vert \beta\right\vert \left[  \left(  2\chi-1\right)  \Phi
^{2}-\chi^{2}\right]  \sin\left(  \varphi+\varphi_{\beta}\right)  \right\}
\text{,}\nonumber\\
\dot{\varphi}  &  =\frac{2}{\left(  \chi-1\right)  \Phi}\left[  \left\vert
\alpha\right\vert \left(  1-\Phi^{2}\right)  \cos\left(  \varphi
-\varphi_{\alpha}\right)  \right.  \left.  +\left\vert \beta\right\vert
\left(  \Phi^{2}-\chi^{2}\right)  \cos\left(  \varphi+\varphi_{\beta}\right)
\right]  +2\left\vert \omega\right\vert \cos\varphi_{\omega} \text{.}
\label{13}%
\end{align}

Here $\varphi_{\alpha}$, $\varphi_{\beta}$ and $\varphi_{\omega}$ are the
polar angles of $\alpha$, $\beta$ and $\omega$, respectively and $\chi$ is a
function of $\Phi$ and $\left\vert z\right\vert $, as defined above.
Therefore, in a similar way to that in Ref. \cite{Musumbu}, we may consider
$\left\vert z\right\vert $ as the only free parameter that determines the
metric, with $\epsilon$ following from the relation
\begin{equation}
\epsilon=\frac{1}{\sqrt{1-\left\vert z\right\vert ^{2}}}\operatorname{arctanh}%
\frac{\sqrt{1-\left\vert z\right\vert ^{2}}\Phi}{\Phi-\left\vert z\right\vert
} =\frac{1}{2\sqrt{1-\left\vert z\right\vert ^{2}}}\ln\left[ \frac{\left(
1+\sqrt{1-\left\vert z\right\vert ^{2}}\right)  \Phi-\left\vert z\right\vert
}{\left(  1-\sqrt{1-\left\vert z\right\vert ^{2}}\right)  \Phi-\left\vert
z\right\vert } \right]  \text{,} \label{14b}%
\end{equation}
as may be derived from the parameter $\Phi=\left\vert z\right\vert /\Gamma
_{-}$, as defined above, which in turn depends, as well as on $\varphi$, also
on the solution of the system (\ref{13}) and the TD coefficients of the
starting Hamiltonian (\ref{5}). Evidently, a given pair ($\left\vert
z\right\vert ,\Phi$), i.e., a given choice of $\left\vert z\right\vert $,
this\ must be further corroborated by a real solution of $\epsilon$ in Eq.
(\ref{14b}), with the argument of the $\operatorname{arctanh}$ ($\ln$) being
not greater than unity (greater than zero), thus demanding $\left\vert
z\right\vert ^{2}>2\Phi/\left(  1+\Phi^{2}\right)  $. We finally observe that
$\left\vert z\right\vert $ can conveniently be considered as a
time-independent parameter, constraining the time-dependence to the remaining
parameters $\varphi$ and $\epsilon$.

\section{Solutions of the Schr{\"o}dinger equation for the generalized
time-dependent Swanson Hamiltonian}

In order to solve the SE for $H(t)$ we shall adapt to the case of TD
non-Hermitian Hamiltonians a method presented in Ref. \cite{MBS} for solving
the SE for TD Hermitian Hamiltonians. This technique takes advantage of a TD
transformation on the SE for the desired Hamiltonian, here a nonunitary
transformation to conform with non-Hermitian Hamiltonians, and the
diagonalization of a TD Invariant within the Lewis and Riesenfeld framework
\cite{LR}. The Lewis and Riesenfeld method ensures that a solution of the SE
governed by a TD Hermitian Hamiltonian $\mathcal{H}(t)$ is an eigenstate of an
associated Hermitian invariant $I(t)$, defined as $\partial_{t}I(t)+i\left[
\mathcal{H}(t),I(t)\right]  =0$, apart from a TD global phase factor. The
method in Ref. \cite{MBS} proposes that, instead of solving the SE for
$\mathcal{H}(t)$ by deriving an invariant directly associated with this
Hamiltonian, a transformation is performed on the SE for bringing the original
Hamiltonian to another form which has already an associated invariant.

The authors in Ref. \cite{MBS} pursued the solution of the SE governed by a
general TD quadratic (Hermitian) Hamiltonian in order to investigate the
mechanism of squeezed states \cite{Walls,Scully} following from the nonlinear
amplification terms of the Hamiltonian. They thus consider the unitary squeeze
operator for transforming the SE for the TD quadratic Hamiltonian, reducing it
to a form associated with a linear Hamiltonian which has already an associated
invariant \cite{PLawande}. Here, we shall focus on the method to approach a TD
non-Hermitian Hamiltonian, leaving open the analysis of the squeezing
mechanism coming from the nonlinear terms of a TD non-Hermitian Hamiltonian.

In the present contribution a similar strategy to that in Ref. \cite{MBS} will
be used, starting from the non-Hermitian $H(t)$ and then deriving the
transformed Hermitian $h(t)$ through the metric operator $\eta(t)$, instead of
a unitary transformation. We further identify this transformed Hamiltonian
with the Hermitian quadratic one treated in Ref. \cite{MBS}, whose solutions
have been derived. Evidently, we must disregard the linear amplification
process considered in Ref. \cite{MBS} since it is absent from $h(t)$. To this
end, we next rewrite the coefficients of the Hermitian (\ref{10}) considering
the Eqs. (\ref{12}) and (\ref{13}). Under the Eqs. (\ref{12}) and (\ref{13})
we obtain the real frequency%
\begin{equation}
W(\left\vert z\right\vert ,\varphi,t)=\left\vert \omega\right\vert \cos
\varphi_{\omega}+\frac{2\Phi}{1-\chi}\left[  \left\vert \alpha\right\vert
\cos\left(  \varphi-\varphi_{\alpha}\right)  -\left\vert \beta\right\vert
\cos\left(  \varphi+\varphi_{\beta}\right)  \right]  \text{.} \label{15}%
\end{equation}
From the system (\ref{13}) we obtain $V(\left\vert z\right\vert ,\varphi
,t)=T^{\ast}(\left\vert z\right\vert ,\varphi,t)=V_{R}(\left\vert z\right\vert
,\varphi,t)+iV_{I}(\left\vert z\right\vert ,\varphi,t)=\kappa(\left\vert
z\right\vert ,\varphi,t)e^{i\zeta(\left\vert z\right\vert ,\varphi,t)}$, with
$\kappa=\left(  V_{R}^{2}+V_{I}^{2}\right)  ^{1/2}$, $\zeta=\arctan\left(
V_{I}/V_{R}\right)  $, and
\begin{subequations}
\label{F}%
\begin{align}
V_{R}(\left\vert z\right\vert ,\varphi,t)  &  =\frac{1}{1-\chi}\left(
\left\vert \omega\right\vert \Phi\sin\varphi_{\omega}\sin\varphi+\left\vert
\alpha\right\vert \cos\varphi_{\alpha}-\left\vert \beta\right\vert \chi
\cos\varphi_{\beta}\right)  \text{,}\label{Fa}\\
V_{I}(\left\vert z\right\vert ,\varphi,t)  &  =\frac{1}{\chi-1}\left(
\left\vert \omega\right\vert \Phi\sin\varphi_{\omega}\cos\varphi-\left\vert
\alpha\right\vert \sin\varphi_{\alpha}-\left\vert \beta\right\vert \chi
\sin\varphi_{\beta}\right)  \text{.} \label{Fb}%
\end{align}
Note that when starting with a Hermitian Hamiltonian (\ref{5}), with real
$\omega$ and $\alpha=\beta^{\ast}$, we verify from Eqs. (\ref{15}) and Eq.
(\ref{F}) that $W(\left\vert z\right\vert ,\varphi,t)=\left\vert
\omega\right\vert $ and $V(\left\vert z\right\vert ,\varphi,t)=\alpha(t)$,
such that $h=H$.

The solutions of the Schr\"{o}dinger equation generated by Hamiltonian
(\ref{10}), given in Ref. \cite{MBS} as
\end{subequations}
\begin{equation}
\left\vert v_{n}(t)\right\rangle =U(t)\left\vert n\right\rangle \text{,}
\label{17}%
\end{equation}
define a complete set of states, $\left\vert n\right\rangle $ being the Fock
states and $U(t)$ the unitary operator
\begin{equation}
U(t)=\Upsilon(t)S\left[  \xi(t)\right]  D\left[  \theta(t)\right]  R\left[
\Omega(t)\right]  \text{.} \label{18}%
\end{equation}
Here $S\left[  \xi(t)\right]  =\exp\left\{  \left[  \xi(t)a^{\dagger2}%
-\xi^{\ast}(t)a^{2}\right]  /2\right\}  $ is the squeeze operator, with
$\xi(t)=r(t)e^{i\phi(t)}$ defining the squeeze parameters, which follow from
another set of coupled nonlinear differential equations
\begin{subequations}
\label{19}%
\begin{align}
\dot{r}(t)  &  =-2\kappa(t)\sin\left[  \zeta(t)-\phi(t)\right]  \text{,}%
\label{19a}\\
\dot{\phi}(t)  &  =-2W(t)-4\kappa(t)\coth\left[  2r(t)\right]  \cos\left[
\zeta(t)-\phi(t)\right]  \text{.} \label{19b}%
\end{align}
where $D\left[  \theta(t)\right]  =\exp\left[  \theta(t)a^{\dagger}%
-\theta^{\ast}(t)a\right]  $ is the displacement operator and $\theta(t)$
satisfies the equation $i\dot{\theta}(t)=\Omega(t)\theta(t)$, with
\end{subequations}
\begin{subequations}
\begin{equation}
\Omega(t)=W(t)+2\kappa(t)\tanh r(t)\cos\left[  \zeta(t)-\phi(t)\right]
\text{.} \label{20}%
\end{equation}
Finally, $R\left[  \Omega(t)\right]  =\exp\left[  -i\varpi(t)a^{\dagger
}a\right]  $ is the rotation operator, with $\varpi(t)=\int_{0}^{t}%
\Omega(t^{\prime})dt^{\prime}$, and $\Upsilon(t)=\exp\left(  -i\varpi
(t)/2\right)  $ is a global phase factor.

Having the wave vectors in Eq. (\ref{17}), we directly obtain the solutions of
the Schr\"{o}dinger equation generated by Hamiltonian (\ref{5}), given by%
\end{subequations}
\begin{equation}
\left\vert \psi_{n}(t)\right\rangle =\eta^{-1}(t)\left\vert v_{n}%
(t)\right\rangle =\eta^{-1}(t)U(t)\left\vert n\right\rangle \text{.}\label{21}%
\end{equation}
For a generic superposition $\left\vert \psi(t)\right\rangle =%
{\textstyle\sum\nolimits_{n}}
c_{n}\left\vert \psi_{n}(t)\right\rangle $ it follows that
\begin{equation}
\left\vert \psi(t)\right\rangle =\eta^{-1}(t)V(t)\left\vert v(0)\right\rangle
\text{,}\label{22}%
\end{equation}
with the evolution operator%
\begin{equation}
V(t)=U(t)U^{\dagger}(0)=\Upsilon(t)S\left[  \xi(t)\right]  D\left[
\theta(t)\right]  R\left[  \Omega(t)\right]  S^{\dagger}\left[  \xi(0)\right]
D^{\dagger}\left[  \theta(0)\right]  \text{,}.\label{23}%
\end{equation}

At this point it is worth mentioning a theorem which can be straightforwardly
adapted from Ref. \cite{MBS} to the context of TD non-Hermitian quantum
mechanics: \textit{If }$I(t)$\textit{ is an invariant associated with an
non-Hermitian Hamiltonian }$H(t)$\textit{, then }$I_{\eta}(t)=\eta
(t)I(t)\eta^{-1}(t)$\textit{ is also an invariant but associated with the
transformed Hermitian Hamiltonian }$h(t)$\textit{, both invariants }%
$I(t)$\textit{ and }$I_{\eta}(t)$\textit{ sharing the same eigenvalue
spectrum. Moreover, the Lewis and Riesenfeld phase is invariant under the
transformation }$\eta(t)$\textit{.} It is not difficult to see that this
theorem fully supports the solutions presented in Eqs. (\ref{21}) and
(\ref{22}).

Before analyzing the observables associated with the pseudo-Hermitian $H(t)$,
it is worth addressing two particular cases: when the coefficients of $H(t)$
are real TD functions and when considering a time-independent metric operator.

\subsection{On the solutions for the TD coupled differential equations
(\ref{13}), (\ref{19}) and (\ref{25})}

Before addressing particular cases where the coefficients of the Hamiltonian
(\ref{5}) are real TD functions and/or a time-independent metric operator is
considered, we add a few comment on the coupled equations ruling the evolution
of the metric parameters $\Phi$ and $\varphi$ [Eqs. (\ref{13}) and (\ref{25})]
and the squeezing parameters $r$ and $\phi$ [Eq. (\ref{19})]. As advanced in
Ref. \cite{MBS}, despite its time dependence, the system (\ref{19}) can be
solved analytically, by quadrature, under particular constraints linking
together its TD functions and thus leaving a lower degree of arbitrariness.
Some solutions for system (\ref{19}) have been presented in Ref. \cite{MBS},
and reasoning by analogy with this reference it will be possible to find
analytical solutions for the systems (\ref{13}) and (\ref{25}), at least for
some specific demands on the TD functions. For example, considering a real TD
function
\begin{equation}
\omega(t)\equiv\frac{\dot{f}(t)}{2}+2\left\vert \beta\right\vert \Phi
\cos\left(  \varphi-\varphi_{\alpha}\right)
\end{equation}
and $\varphi_{\beta}(t)=-\varphi_{\alpha}(t)$, we eliminate the parameter time
from the system (\ref{13}), to obtain, with $\varsigma(t)=\varphi(t)+f(t)$ and
a constant $v=$ $\varphi_{\alpha}(t)+f(t)$, the first order differential
equation%
\begin{equation}
\frac{d\Phi}{d\varsigma}=\frac{\Phi}{\tan\left(  \varsigma-v\right)  }\text{,}%
\end{equation}
whose integration leads to a constant of motion and thus to the solutions for
$\Phi$ and $\varphi$.

\section{Particular cases}

\subsection{The generalized TD Swanson's Hamiltonian with real coefficients
$\omega(t),\alpha(t),\beta(t)$}

When considering the TD coefficients $\omega(t),\alpha(t),\beta(t)$ to be real
functions instead of complex ones, the equations in Sections III and IV
considerably simplify. Let us start by demanding $h(t)$ in Eq. (\ref{10}) to
be Hermitian. By imposing $W$ to be real we now obtain%
\begin{equation}
\dot{\lambda}_{0}=2\Phi\left[  \dot{\Phi}+2\left(  \alpha-\beta\chi\right)
\sin\varphi\right]  \text{,} \label{24}%
\end{equation}
while the imposition $T=V^{\ast}$ leads to the simplified coupled nonlinear
differential equations
\begin{subequations}
\label{25}%
\begin{align}
\dot{\Phi}  &  =\frac{2}{\chi-1}\left\{  \alpha\left(  1-\Phi^{2}\right)
+\beta\left[  \left(  2\chi-1\right)  \Phi^{2}-\chi^{2}\right]  \right\}
\sin\varphi\text{,}\label{25a}\\
\dot{\varphi}  &  =2\omega-\frac{2}{\left(  1-\chi\right)  \Phi}\left[
\alpha\left(  1-\Phi^{2}\right)  +\beta\left(  \Phi^{2}-\chi^{2}\right)
\right]  \cos\varphi\text{.} \label{25b}%
\end{align}
Again, $\left\vert z\right\vert $ can be taken as the only free parameter that
determines the metric, with $\epsilon$ following from Eq. (\ref{14b}). To
further identify the transformed Hermitian $h(t)$ with the quadratic
Hamiltonian whose SE is solved in Ref. \cite{MBS}, we rewrite the coefficients
of $h(t)$ considering the Eqs. (\ref{24}) and (\ref{25}). We thus obtain the
real frequency%
\end{subequations}
\begin{equation}
W(\left\vert z\right\vert ,\varphi,t)=\omega+\frac{2\Phi}{1-\chi}\left[
\alpha-\beta\right]  \cos\varphi\text{,} \label{26}%
\end{equation}
and the simplified real function
\begin{subequations}
\label{16}%
\begin{equation}
V(\left\vert z\right\vert ,\varphi,t)=T(\left\vert z\right\vert ,\varphi
,t)=\kappa(\left\vert z\right\vert ,\varphi,t)=\frac{\alpha-\beta\chi}{1-\chi
}\text{.} \label{27}%
\end{equation}
From the above equations the solutions presented in Eqs. (\ref{21}) and
(\ref{22}) follow straightforwardly.

\subsection{The generalized TD Swanson's Hamiltonian with a time-independent
metric operator}

When a time-independent metric operator is considered, the coefficients of the
transformed Hamiltonian $h(z,\epsilon,t)$ simplify to
\end{subequations}
\begin{subequations}
\label{28}%
\begin{align}
W(z,\epsilon,t) &  =-\frac{1}{\lambda_{0}}\left[  \omega\left(  \chi
+\lambda_{+}\lambda_{-}\right)  +2\left(  \alpha\lambda_{+}+\beta\chi
\lambda_{-}\right)  \right]  \text{,}\label{28a}\\
V(z,\epsilon,t) &  =\frac{1}{\lambda_{0}}\left(  \alpha+\omega\lambda
_{-}+\beta\lambda_{-}^{2}\right)  \text{,}\label{28b}\\
T(z,\epsilon,t) &  =\frac{1}{\lambda_{0}}\left(  \omega\chi\lambda_{+}%
+\alpha\lambda_{+}^{2}+\beta\chi^{2}\right)  \text{,}\label{28c}%
\end{align}
For $h$ to be Hermitian we again impose $W$ to be real and $T=V^{\ast}$. The
first constraint leads to the relation
\end{subequations}
\begin{equation}
\left\vert \omega\right\vert \left(  \chi+\Phi^{2}\right)  \sin\varphi
_{\omega}+2\Phi\left[  \left\vert \alpha\right\vert \sin\left(  \varphi
-\varphi_{\alpha}\right)  -\left\vert \beta\right\vert \chi\sin\left(
\varphi+\varphi_{\beta}\right)  \right]  =0\text{,}\label{29}%
\end{equation}
while the latter gives rise to the equations
\begin{subequations}
\label{30}%
\begin{align}
\left\vert \omega\right\vert \left(  1-\chi\right)  \Phi\cos\varphi_{\omega
}-\left\vert \alpha\right\vert \left(  1-\Phi^{2}\right)  \cos\left(
\varphi-\varphi_{\alpha}\right)  +\left\vert \beta\right\vert \left(  \chi
^{2}-\Phi^{2}\right)  \cos\left(  \varphi+\varphi_{\beta}\right)   &
=0\text{,}\label{30a}\\
\left\vert \omega\right\vert \left(  1+\chi\right)  \Phi\sin\varphi_{\omega
}+\left\vert \alpha\right\vert \left(  1+\Phi^{2}\right)  \sin\left(
\varphi-\varphi_{\alpha}\right)  -\left\vert \beta\right\vert \left(  \chi
^{2}+\Phi^{2}\right)  \sin\left(  \varphi+\varphi_{\beta}\right)   &
=0\text{.}\label{30b}%
\end{align}
From Eqs. (\ref{29}) and (\ref{30b}) we obtain the relation%
\end{subequations}
\begin{equation}
\left\vert \alpha\right\vert \left(  1-\Phi^{2}\right)  \sin\left(
\varphi-\varphi_{\alpha}\right)  =\left\vert \beta\right\vert \left(  \chi
^{2}-\Phi^{2}\right)  \sin\left(  \varphi+\varphi_{\beta}\right)
\text{,}\label{31}%
\end{equation}
which, together with Eq. (\ref{30a}), gives us
\begin{subequations}
\label{32}%
\begin{align}
\sin\left(  \varphi-\varphi_{\alpha}\right)   &  =\frac{\left\vert
\beta\right\vert \left(  \chi^{2}-\Phi^{2}\right)  }{\left\vert \omega
\right\vert \left(  1-\chi\right)  \Phi\cos\varphi_{\omega}}\sin\left(
\varphi_{\alpha}+\varphi_{\beta}\right)  \text{,}\label{32a}\\
\sin\left(  \varphi+\varphi_{\beta}\right)   &  =\frac{\left\vert
\alpha\right\vert \left(  1-\Phi^{2}\right)  }{\left\vert \omega\right\vert
\left(  1-\chi\right)  \Phi\cos\varphi_{\omega}}\sin\left(  \varphi_{\alpha
}+\varphi_{\beta}\right)  \text{.}\label{32b}%
\end{align}
By substituting Eq. (\ref{32}) back into Eq. (\ref{29}), we finally obtain the
equation%
\end{subequations}
\begin{equation}
\left\vert z\right\vert \Phi^{3}+\left(  2-\left\vert z\right\vert
^{2}\right)  \Phi^{2}-3\left\vert z\right\vert \Phi+\left\vert z\right\vert
^{2}=0\text{,}\label{33}%
\end{equation}
whose roots enable us to compute $\varphi$ from Eq. (\ref{32}) and then
$\epsilon$ from the relation given in Eq. (\ref{14b}). Here, the real
frequency $W(\left\vert z\right\vert ,\varphi,t)$ and the complex
$V(\left\vert z\right\vert ,\varphi,t)=T^{\ast}(\left\vert z\right\vert
,\varphi,t)$ still follow from Eqs. (\ref{15}) and (\ref{F}), respectively,
with time-independent $z$ and $\epsilon$.

\subsubsection{A time-independent metric operator with real TD coefficients
$\omega(t),\alpha(t),\beta(t)$}

When a time-independent metric operator is considered together with real TD
parameters $\omega(t),\alpha(t),\beta(t)$, it follows from Eq. (\ref{29}) that
$\varphi=0$ and from Eq. (\ref{30a}) we derive the equation
\begin{equation}
\left(  \left\vert \alpha\right\vert -\left\vert \beta\right\vert \right)
\Phi^{2}+\left\vert \omega\right\vert \left(  1-\chi\right)  \Phi-\left\vert
\alpha\right\vert +\left\vert \beta\right\vert \chi^{2}=0 \label{34}%
\end{equation}
which leads to the relation%
\begin{equation}
\frac{\tanh(2\Xi)}{\tilde{\Xi}}=\frac{\alpha-\beta}{\alpha+\beta-z\omega
}\text{,} \label{35}%
\end{equation}
and, consequently, to the metric parameter
\begin{subequations}
\label{36}%
\begin{align}
\epsilon &  =\frac{1}{2\sqrt{1-\left\vert z\right\vert ^{2}}}%
\operatorname{arctanh}\frac{\left(  \left\vert \alpha\right\vert -\left\vert
\beta\right\vert \right)  \sqrt{1-\left\vert z\right\vert ^{2}}}{\left\vert
\alpha\right\vert +\left\vert \beta\right\vert -\left\vert z\right\vert
\left\vert \omega\right\vert }\label{36a}\\
&  =\frac{1}{4\sqrt{1-\left\vert z\right\vert ^{2}}}\ln\frac{\left\vert
\alpha\right\vert +\left\vert \beta\right\vert -\left\vert z\right\vert
\left\vert \omega\right\vert +\left(  \left\vert \alpha\right\vert -\left\vert
\beta\right\vert \right)  \sqrt{1-\left\vert z\right\vert ^{2}}}{\left\vert
\alpha\right\vert +\left\vert \beta\right\vert -\left\vert z\right\vert
\left\vert \omega\right\vert -\left(  \left\vert \alpha\right\vert -\left\vert
\beta\right\vert \right)  \sqrt{1-\left\vert z\right\vert ^{2}}}\text{.}
\label{36b}%
\end{align}
However, a time-independent metric brings about the constraint on the TD
parameters of the Hamiltonian
\end{subequations}
\begin{equation}
\frac{\left\vert \dot{\alpha}\right\vert +\left\vert \dot{\beta}\right\vert
-\left\vert z\right\vert \left\vert \dot{\omega}\right\vert +\left(
\left\vert \dot{\alpha}\right\vert -\left\vert \dot{\beta}\right\vert \right)
\sqrt{1-\left\vert z\right\vert ^{2}}}{\left\vert \alpha\right\vert
+\left\vert \beta\right\vert -\left\vert z\right\vert \left\vert
\omega\right\vert +\left(  \left\vert \alpha\right\vert -\left\vert
\beta\right\vert \right)  \sqrt{1-\left\vert z\right\vert ^{2}}}%
=\frac{\left\vert \dot{\alpha}\right\vert +\left\vert \dot{\beta}\right\vert
-\left\vert z\right\vert \left\vert \dot{\omega}\right\vert -\left(
\left\vert \dot{\alpha}\right\vert -\left\vert \dot{\beta}\right\vert \right)
\sqrt{1-\left\vert z\right\vert ^{2}}}{\left\vert \alpha\right\vert
+\left\vert \beta\right\vert -\left\vert z\right\vert \left\vert
\omega\right\vert -\left(  \left\vert \alpha\right\vert -\left\vert
\beta\right\vert \right)  \sqrt{1-\left\vert z\right\vert ^{2}}}\text{,}
\label{37}%
\end{equation}
where we have assume a time-independent $\left\vert z\right\vert $ as the only
free parameter that determines the metric, with $\epsilon$ following from Eq.
(\ref{38}). The existence of a real solution for $\epsilon$ demands the
argument of the $\operatorname{arctanh}$ ($\ln$) to be not greater than unity
(to be greater than zero), and consequently, there is no real solution for
$\left\vert z\right\vert \in\left[  \left\vert z_{-}\right\vert ,\left\vert
z_{+}\right\vert \right]  $, with%

\begin{equation}
\left\vert z_{\pm}\right\vert =\frac{\left(  \left\vert \alpha\right\vert
+\left\vert \beta\right\vert \right)  \left\vert \omega\right\vert \pm\left(
\left\vert \alpha\right\vert -\left\vert \beta\right\vert \right)  \left(
\left\vert \omega\right\vert ^{2}-4\left\vert \alpha\right\vert \left\vert
\beta\right\vert \right)  }{\left\vert \omega\right\vert ^{2}+\left(
\left\vert \alpha\right\vert -\left\vert \beta\right\vert \right)  ^{2}%
}\text{.} \label{38}%
\end{equation}
The roots $\left\vert z_{\pm}\right\vert $ present the same form as those in
Ref. \cite{Musumbu}, the difference here being that $\left\vert \omega
\right\vert $, $\left\vert \alpha\right\vert $, and $\left\vert \beta
\right\vert $ are TD functions instead of constant parameters, additionally
constrained by Eq. (\ref{37}), thus placing an additional difficulty for the
observance of the requirements for a real solution for $\epsilon$. Finally,
when we identify the Hamiltonian $h(z,\epsilon,t)$ with the Hermitian
quadratic one in (\cite{MBS}) we obtain for $W(\left\vert z\right\vert
,\varphi,t)$ and $V(\left\vert z\right\vert ,\varphi,t)=T^{\ast}(\left\vert
z\right\vert ,\varphi,t)=\kappa(\left\vert z\right\vert ,\varphi
,t)e^{i\zeta(\left\vert z\right\vert ,\varphi,t)}$, the same expressions as in
Eqs. (\ref{26}) and (\ref{27}), respectively, with time-independent
$\left\vert z\right\vert $ and $\epsilon$.

\section{Observables}

\subsection{The generalized TD Swanson Hamiltonian}

Considering the observables for the generalized TD Swanson Hamiltonian, we
start by focusing on the derivation of all the Hermitian operators on the
continuous variety of Hilbert spaces $\mathcal{H}_{z}$ for any $|z|\in
\lbrack-1,1]$. As argued in \cite{Andreas}, the Hamiltonian $H$ itself is not
one of the Hermitian operator due the presence of the gauge-like term in Eq.
(\ref{1}). Using Eq. (\ref{14b}) to rewrite the metric operator in Eq.
(\ref{6}) in the form \cite{Musumbu,GHS}
\begin{subequations}
\label{39}%
\begin{align}
\eta(t)  &  =\left(  \frac{\left(  1+\sqrt{1-\left\vert z\right\vert ^{2}%
}\right)  \Phi-\left\vert z\right\vert }{\left(  1-\sqrt{1-\left\vert
z\right\vert ^{2}}\right)  \Phi-\left\vert z\right\vert }\right)
^{\frac{a^{\dagger}a+\frac{1}{2}\left(  za^{2}+z^{\ast}a^{\dagger2}\right)
+\frac{1}{2}}{2\sqrt{1-\left\vert z\right\vert ^{2}}}}\label{39a}\\
&  =\left(  \frac{\left(  1+\sqrt{1-\left\vert z\right\vert ^{2}}\right)
\Phi-\left\vert z\right\vert }{\left(  1-\sqrt{1-\left\vert z\right\vert ^{2}%
}\right)  \Phi-\left\vert z\right\vert }\right)  ^{\frac{\left(  1-\left\vert
z\right\vert \cos\varphi\right)  p^{2}+\left(  1+\left\vert z\right\vert
\cos\varphi\right)  \omega^{2}x^{2}-\left\vert z\right\vert \omega\sin
\varphi\left\{  x,p\right\}  }{4\omega\sqrt{1-\left\vert z\right\vert ^{2}}}}
, \label{39b}%
\end{align}
which we use to solve the quasi-Hermiticity condition $O^{\dagger}%
(t)\mu(t)=\mu(t)O(t)$. Given (\ref{39}), we only find the observables%
\end{subequations}
\begin{equation}
O(t)=\left(  1-\left\vert z\right\vert \cos\varphi\right)  p^{2}+\left(
1+\left\vert z\right\vert \cos\varphi\right)  \omega^{2}x^{2}-\left\vert
z\right\vert \omega\sin\varphi\left\{  x,p\right\}  \text{,} \label{40}%
\end{equation}
demonstrating that neither the position $x=\frac{1}{\sqrt{2\omega}}\left(
a+a^{\dagger}\right)  $ nor the momentum $p=i\sqrt{\frac{\omega}{2}}\left(
a^{\dagger}-a\right)  $ operators remain Hermitian as they are in the standard
$L^{2}$-metric, with regard to the TD $\eta(t)$-metric even for particular
choices of $\left\vert z\right\vert $. Using the relation $O(t)=\eta
^{-1}(t)o\eta(t)$ together with Eq. (\ref{13}) we may compute the
quasi-Hermitian position $X(t)$ and momentum $P(t)$ operators
\begin{subequations}
\label{41}%
\begin{align}
X(t)  &  =\frac{1}{\left\vert z\right\vert \sqrt{\Phi^{2}-\chi}}\left\{
\left[  \left(  1-i\left\vert z\right\vert \sin\varphi\right)  \Phi-\left\vert
z\right\vert \right]  x+\frac{i}{\omega}\left(  1-\left\vert z\right\vert
\cos\varphi\right)  \Phi p\right\}  \text{,}\label{41a}\\
P(t)  &  =\frac{1}{\left\vert z\right\vert \sqrt{\Phi^{2}-\chi}}\left\{
\left[  \left(  1+i\left\vert z\right\vert \sin\varphi\right)  \Phi-\left\vert
z\right\vert \right]  p-i\omega\left(  1+\left\vert z\right\vert \cos
\varphi\right)  \Phi x\right\}  \text{,} \label{41b}%
\end{align}
corroborating the conclusion we have drawn from Eq. (\ref{40}).

\subsection{Particular cases}

The observables computed above in Eqs. (\ref{40}) and (\ref{41}) also apply to
the cases where real coefficients $\omega(t),\alpha(t),\beta(t)$ are assumed
and when a time-independent metric operator is considered, the difference
being that $\Phi$ and $\varphi$ now follow, instead of Eq. (\ref{13}), from
the coupled Eqs. (\ref{25}) in the former case, and from Eqs. (\ref{32}) and
(\ref{33}) in the latter case. However, when a time-independent metric
operator is considered simultaneously with real coefficients $\omega
(t),\alpha(t),\beta(t)$, the Hermitian observables in Eq. (\ref{40}) and those
in Eq. (\ref{41}) simplify to
\end{subequations}
\begin{subequations}
\label{42}%
\begin{align}
O(t)  &  =\left(  1-\left\vert z\right\vert \right)  p^{2}+\left(
1+\left\vert z\right\vert \right)  \omega^{2}x^{2}\text{,}\label{42a}\\
X(t)  &  =\frac{1}{\left\vert z\right\vert \sqrt{\Phi^{2}-\chi}}\left[
\left(  \Phi-\left\vert z\right\vert \right)  x+\frac{i}{\omega}\left(
1-\left\vert z\right\vert \right)  \Phi p\right]  \text{,}\nonumber\\
&  =\cosh\left(  \Xi\right)  x+\frac{i}{\omega}\frac{\left(  1-\left\vert
z\right\vert \right)  }{\sqrt{1-\left\vert z\right\vert ^{2}}}\sinh\left(
\Xi\right)  p\text{,}\label{42b}\\
P(t)  &  =\frac{1}{\left\vert z\right\vert \sqrt{\Phi^{2}-\chi}}\left[
\left(  \Phi-\left\vert z\right\vert \right)  p-i\omega\left(  1+\left\vert
z\right\vert \right)  \Phi x\right]  \text{,}\nonumber\\
&  =\cosh\left(  \Xi\right)  p-i\omega\frac{\left(  1+\left\vert z\right\vert
\right)  }{\sqrt{1-\left\vert z\right\vert ^{2}}}\sinh\left(  \Xi\right)
x\text{.} \label{42c}%
\end{align}
The Eqs. (\ref{42}) are exactly of the same form as those in Ref.
\cite{Musumbu}, the difference being that here we have TD parameters.
Therefore, when considering the Hamiltonian (\ref{5}) with time-independent
real parameters together with a time-independent metric operator, it is
straightforward to verify that all the above derivations are in complete
agreement with those in \cite{Musumbu}.

\section{Conclusion}

We have studied a generalized Swanson Hamiltonian allowing for TD complex
coefficients and a TD metric operator. We treated the model within the
framework introduced in Ref. \cite{Andreas} where, despite the lack of the
observability of the non-Hermitian Hamiltonian under a TD metric operator,
their associated observables are computed as in the case where a
time-independent metric is considered. To solve the SE for the generalized TD
Swanson's Hamiltonian we have adapted a technique presented in Ref. \cite{MBS}
which\ relies on the Lewis and Riesenfeld TD invariants. Apart from deriving
the solutions of the SE for our TD non-Hermitian Hamiltonian we have thus
computed their associated observables, analyzing particular cases where a
time--independent metric operator is considered and TD real coefficients are
assumed for the non-Hermitian Hamiltonian.

From the results presented here we may next explore some interesting
applications such as the generation of squeezing from a non-Hermitian
parametric oscillator. Moreover, our TD Hamiltonian can be also considered to
describe the non-Hermitian dynamical Casimir effect, and thus the rate of
particles creations resulting, for example, from the accelerated movement of a
cavity mirror can also be computed. The results for the generation of
squeezing and the rate of photon creation derived from a non-Hermitian
quadratic Hamiltonian can then be compared with the well-known results coming
from the Hermitian Hamiltonians, thus delivering more timely hints on the
physics of non-Hermitian Hamiltonians.

As another application motivated by this work is the possibility of
engineering effective non-Hermitian Hamiltonians within trapped ions, circuit
or cavity QED, NMR and other systems presenting great flexibility of handling
its internal interactions. By mastering not only the technique for treating
non-Hermitian Hamiltonians, but also for constructing non-Hermitian
interactions, we may seek to contribute with the implementation of processes
such as quantum simulation and quantum logical implementation, bringing
additional ingredients to the usual Hermitian quantum mechanics.
\end{subequations}
\begin{flushleft}
{\Large \textbf{Acknowledgements}}
\end{flushleft}

M.H.Y. Moussa wishes to express his thanks to CAPES, Brazilian financial
agency, and City University London for kind hospitality.

\end{document}